\documentclass[epj]{svjour}
\usepackage{times}
\usepackage{graphicx}
\newfont{\eufm}{eufm10}
\newfont{\cmu}{cmu10}

\begin{document}

\title{Beta-decay branching ratios of $^{62}$Ga}

\author{
A. Bey\inst{1} \and
B. Blank\inst{1} \and
G. Canchel\inst{1} \and
C. Dossat\inst{1} \and
J. Giovinazzo\inst{1} \and
I. Matea\inst{1} \and
V.-V. Elomaa\inst{2} \and
T. Eronen\inst{2} \and
U. Hager\inst{2} \footnote[1] \and ,
J. Hakala\inst{2} \and
A. Jokinen\inst{2} \and
A. Kankainen\inst{2} \and
I. Moore\inst{2} \and
H. Penttil{\"a}\inst{2} \and
S. Rinta-Antila\inst{2} \footnote[2] \and ,
A. Saastamoinen\inst{2} \and
T. Sonoda\inst{2} \footnote[3] \and ,
J. {\"A}yst{\"o}\inst{2} \and
N. Adimi\inst{3} \and
G. de France\inst{4} \and
J.-C. Thomas\inst{4} \and
G. Voltolini\inst{4} \and
T. Chaventr\'e\inst{5}
}

\institute{Centre d'Etudes Nucl\'eaires de Bordeaux Gradignan -
Universit\'e Bordeaux 1 - UMR 5797 CNRS/IN2P3, Chemin du Solarium,
BP 120, 33175~Gradignan Cedex, France
\and
Department of Physics, University of Jyv\"askyl\"a,
P.O. Box 35, FIN-40351 Jyv\"askyl\"a, Finland
\and
Facult\'e de Physique, USTHB, B.P.32, El Alia, 16111 Bab Ezzouar, Alger, Algeria
\and
Grand Acc\'el\'erateur National d'Ions Lourds, CEA/DSM - CNRS/IN2P3, B.P. 55027, F-14076 Caen Cedex 5, France
\and
Laboratoire de Physique Corpusculaire, IN2P3-CNRS, ISMRA et Universit{\'e} de Caen, 6 bvd Mar{\'e}chal Juin,
14050 Caen Cedex, France
}

\abstract{
Beta-decay branching ratios of $^{62}$Ga have been measured at the IGISOL facility of the Accelerator Laboratory
of the University of Jyv{\"a}skyl{\"a}. $^{62}$Ga is one of the heavier T$_z$ = 0, $0^+ \rightarrow 0^+$ 
$\beta$-emitting nuclides used to determine the vector coupling constant of the weak interaction and 
the $V_{ud}$ quark-mixing matrix element. For part of the experimental studies presented here, the JYFLTRAP
facility has been employed to prepare isotopically pure beams of $^{62}$Ga. The branching ratio obtained, 
$BR$~= 99.893(24)~\%, for the super-allowed branch is in agreement with previous measurements and allows 
to determine the $ft$ value and the universal $\mathcal{F}t$ value for the super-allowed $\beta$ decay of $^{62}$Ga.
}


\maketitle

\renewcommand{\thefootnote}{\alph{footnote}}

\footnotetext[1]{present address: TRIUMF, 4004 Wesbrook Mall, Vancouver, British 
Columbia, V6T 2A3, Canada}
\footnotetext[2]{present address: Department of Physics, Oliver Lodge 
Laboratory, University of Liverpool, Liverpool L69 7ZE, United Kingdom}
\footnotetext[3]{present address: Instituut voor Kern- en Stralingsfysica, 
Celestijnenlaan 200D, B-3001 Leuven, Belgium}

\section{Introduction}

Nuclear $\beta$ decay is a commonly used probe to study the properties of the atomic nucleus.
As $\beta$ decay is governed by the weak interaction, it may also be used to test the
light-quark sector of the Standard Model (SM).
The SM incorporates the conserved-vector-current (CVC) hypothesis, which assumes that the vector part
of the weak interaction is not influenced by the strong interaction.  Thus, the vector current should
not be renormalized in the nuclear medium.  The comparative half-life $ft$ of a particular class
of $\beta$-decaying nuclides gives access to the vector coupling constant $g_v$ used to test CVC~\cite{hardy05}.
As a further test, the combination of $g_v$ with the muonic vector coupling constant $g_\mu$ allows to determine
the up-quark down-quark element $V_{ud}$ of the Cabibbo-Kobayashi-Maskawa quark-mixing matrix, which in the SM
is unitary.  $V_{ud}$ has by far the most significant weight in such a unitarity test.

Due to their intrinsic simplicity, super-allowed 0$^+ \rightarrow $ 0$^+$
$\beta$ decays (so-called pure Fermi transitions) are the preferred choice to determine
the corrected $\mathcal{F}t $ values:
$$
\mathcal{F}t = ft \times (1-\delta_C + \delta_{NS}) \times (1 + \delta'_R) =
$$
$$
   \;\; \frac{k}{g_v^2 \times \langle M_F \rangle ^2 \times (1 + \Delta_R)}
$$
where $k$ is a product of constants and $\langle M_F \rangle$ is the Fermi-decay matrix element
$ \langle M_F \rangle^2 = T(T+1) - T_{zi}T_{zf}$. $T$ is the isospin of the decaying nucleus
and $T_{zi}$ and $T_{zf}$ are the third
components of $T$ for the initial and final state, respectively.

The experimental quantities necessary for the determination of $ft$ are:  the $\beta$ decay energy $Q_{EC}$,
the half-life $T_{1/2}$, and the super-allowed branching ratio $BR$.
The theoretical corrections $\delta_C$, $\delta_{NS}$, $\delta'_R$ and $\Delta_R$ must
be determined by models~\cite{towner02,towner07}.
$\mathcal{F}t $ values have been determined for thirteen such super-allowed Fermi decays
with a precision close to or better than $10^{-3}$~\cite{hardy05}.

The latest determination of $\mathcal{F}t $ yields an average value of $\mathcal{F}t $ = (3071.4$\pm$0.8)~s~\cite{towner07}.
With the coupling constant for the purely leptonic muon decay, one determines
$V_{ud} = $ 0.97418(26).  Nuclear $\beta$ decay provides the most
precise determination of this matrix element, which dominates the unitarity test.

The main uncertainty in the determination of $\mathcal{F}t $ is due to the uncertainty in the nuclear structure
dependent corrections $\delta_C - \delta_{NS}$, whereas the main uncertainty for the value of $V_{ud}$ is due to the
nucleus independent radiative correction $\Delta_R$. Therefore,
significant progress in this field demands improvements of these theoretical corrections. The radiative correction
$\delta'_R$ could also be improved by adding more terms in its evaluation in the framework of quantum
electrodynamics.

However, experimental data can still test the nuclear structure corrections $\delta_C - \delta_{NS}$.
The experimental $ft$ values corrected only with $(1 + \delta'_R)$ scatter significantly.
When the nuclear structure dependent term $\delta_C - \delta_{NS}$
is added, a constant $\mathcal{F}t $ value is found,
thus verifying the CVC hypothesis and allowing for the determination of
$g_v$ and finally $V_{ud}$.

This test is particularly sensitive for large $\delta_C - \delta_{NS}$ corrections, which
is the case for heavy $T_z$~= 0 nuclei such as $^{62}$Ga.
In this work, we present a precision study of the $\beta$-decay branching ratios of this nucleus.
Several $\beta$-decay branching ratio measurements for $^{62}$Ga have already been 
published~\cite{blank02ga62,doering02,hyman03,canchel05,hyland06}. However, only the last reference~\cite{hyland06} was able to observe 
and give branching ratios for transitions other than the 2$^+$ to 0$^+$ $\gamma$ transition in $^{62}$Zn
and therefore to determine the $\beta$-decay feeding of several $I^\pi$~= 0$^+$ and 1$^+$ states.
The other quantities needed to determine the $ft$ value, i.e. the $\beta$-decay Q value and the half-life, have been 
measured recently with high precision~\cite{hyman03,canchel05,blank04ga62,hyland05,eronen06}.

\section{Experimental procedure}

The experiment was performed at the IGISOL facility in the Accelerator Laboratory of the University of Jyv{\"a}skyl{\"a}.
An intense proton beam (up to 45$\mu$A) at an energy of 48~MeV was directed onto a $^{64}$Zn target of thickness 3~mg/cm$^2$.
The fusion-evaporation residues recoiling out of the target were thermalised in a helium-filled chamber. A gas flow 
extracted the activity out of the target chamber. The singly-charged ions were then accelerated  
to about 30 keV and mass analysed by a dipole magnet with a resolution of 
m/$\Delta$m~$\approx$ 300 and sent to the experimental setup.

\begin{figure}
\begin{center}
\resizebox{.48\textwidth}{!}{\includegraphics[angle=-90]{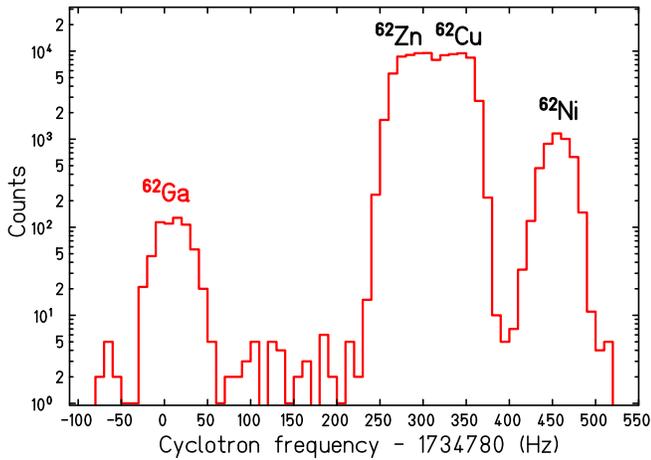}}
\caption[]{Isobaric scan for A=62 with the purification trap of JYFLTRAP. The isobaric components are labeled and illustrated 
           relative to the centre cyclotron frequency of $^{62}$Ga (1734780 Hz). The trap multi-channel plate detector is 
           saturated for $^{62}$Zn and $^{62}$Cu.
}
\label{fig:isobars}
\end{center}
\end{figure}

We used two different schemes to perform the measurements: i) using the 
JYFLTRAP setup to separate $^{62}$Ga from the other A=62 isobars. The JYFLTRAP 
consists of a radiofrequency quadrupole (RFQ) cooler \cite{nieminen02} and of two Penning 
traps. In this case only the first trap, the purification trap~\cite{kolhinen04}, was used
to separate $^{62}$Ga from contaminants, which were mainly $^{62}$Zn and $^{62}$Cu having yields 
of more than 1000 times the yield of $^{62}$Ga. The purification cycle was chosen to be as short as possible, 
in this case 71 ms, limited by the trap cleaning process employing the buffer-gas 
cooling technique~\cite{savard91}. This cycle time additionally sets the requirement for the accumulation time of A=62 ions in the RFQ. 
The isobaric cleaning scan for A=62 is shown in figure~\ref{fig:isobars}. To prepare clean $^{62}$Ga samples, the cyclotron frequency was 
fixed to 1734780 Hz. The cleaned bunch was then ejected from the trap to a movable tape for decay studies.
The collection tape (see below) was moved after 100 to 9000 of such cycles (this parameter was 
changed several times without any significant influence on the data).
The $^{62}$Ga rate was about 100 pps during beam-on periods.

\noindent
ii) using a setup situated directly downstream from the IGISOL focal plane. In this scheme, 
we used two different measurement cycles. In the first, a grow-in time 
of 250~ms was followed by a decay period of 250~ms. The second had only a
grow-in period of 390~ms. For both cycles, the grow-in was preceded by 10~ms 
beam-off time for background determination. At the end of the cycle, the collection tape was moved.
The $^{62}$Ga detection rate varied between about 50 and 120 pps during beam-on periods.

\begin{figure}
\begin{center}
\resizebox{.48\textwidth}{!}{\includegraphics{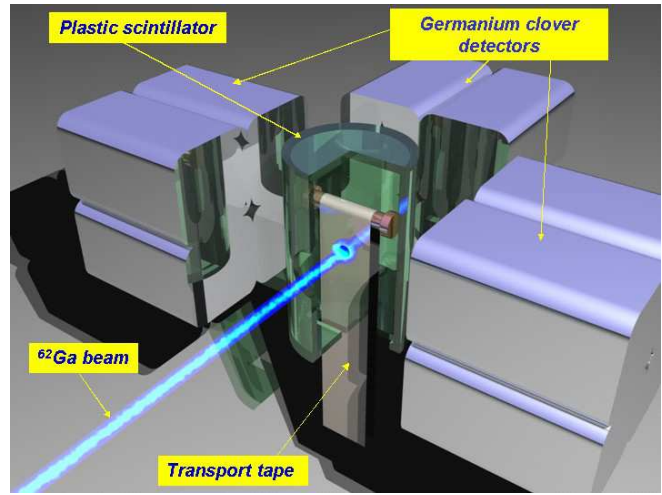}}
\caption[]{Schematic drawing of the experimental set-up used in the present experiment.
           The activity is collected on a tape inside the 4$\pi$ plastic scintillator (entrance hole diameter of 12~mm) 
           viewed by two photomultipliers mounted on top of the scintillator (not shown) to detect $\beta$ particles.
           The scintillator is surrounded in close geometry by three EUROBALL HPGe clover detectors
           for $\gamma$ detection.
}
\label{fig:setup}
\end{center}
\end{figure}

\begin{figure*}[htb]
\begin{center}
\begin{minipage}{16cm}
\includegraphics[scale=0.25]{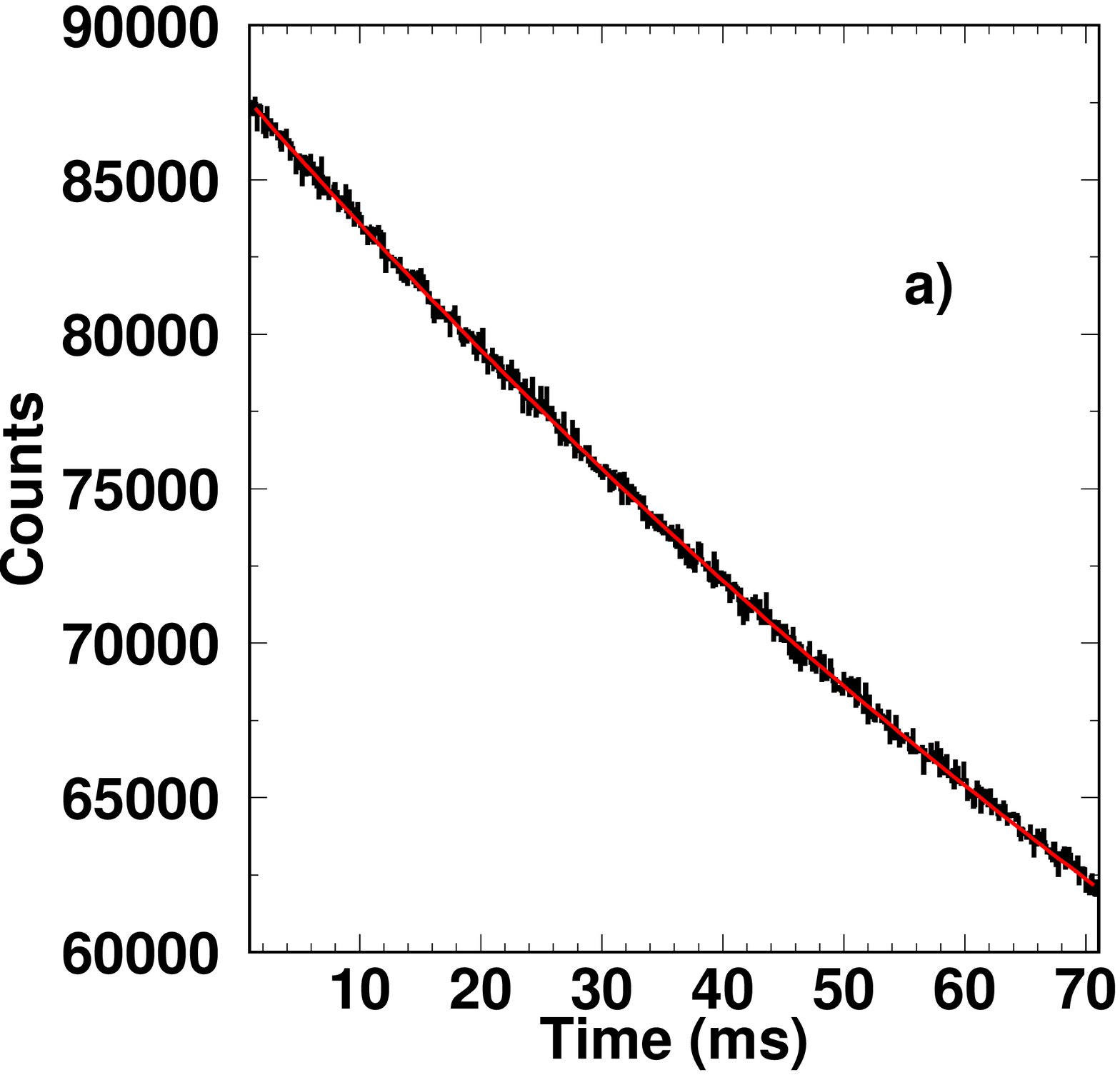} \ \hfill \
\includegraphics[scale=0.25]{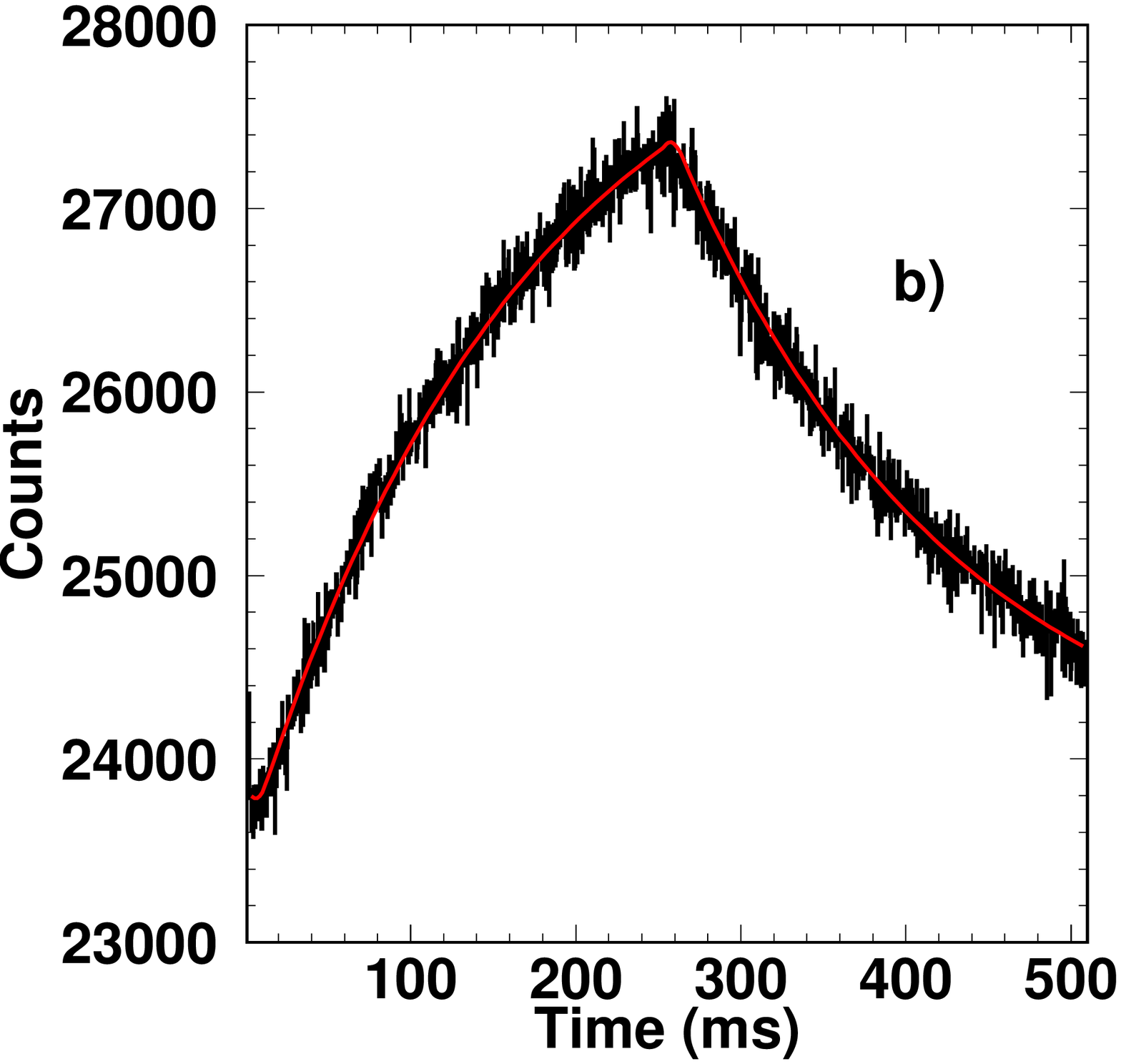} \ \hfill \ 
\includegraphics[scale=0.25]{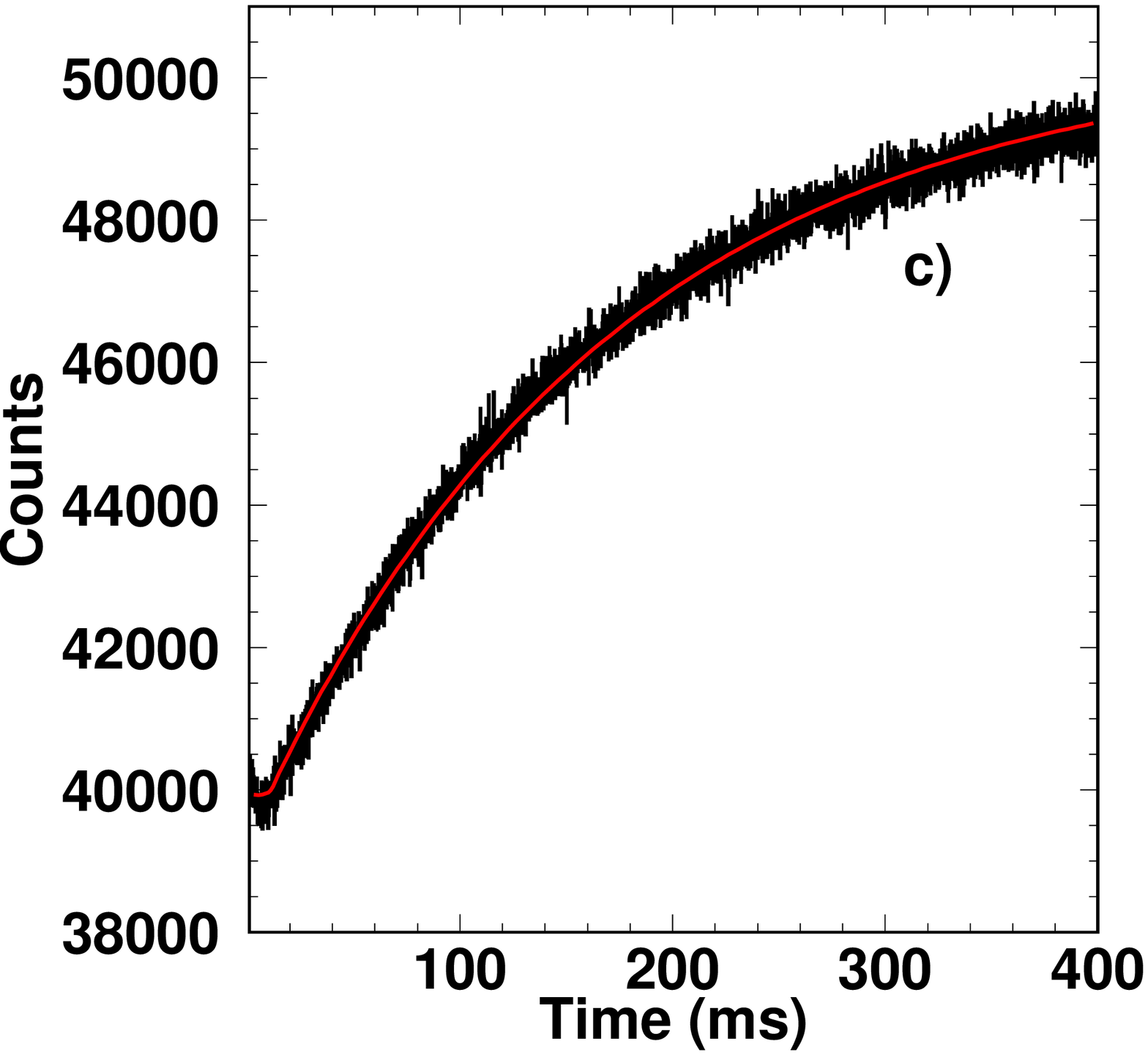} 
\end{minipage}
\caption[]{Beta-decay time distributions determined by means of the plastic scintillator for the three different
           measurement schemes: a) trap-assisted spectroscopy using JYFLTRAP to prepare the pure $^{62}$Ga sample.
           In this case only the decay component of $^{62}$Ga is observed together with a constant background.
           The daughter activity is too weak and too long-lived to be observed. b) Time spectrum from the central 
           IGISOL beam line, where a grow-in period of 250~ms was followed by a 250~ms decay time. The strongly produced
           contaminants $^{62}$Zn and $^{62}$Cu contribute to the constant background. c) After a 390 ms grow-in period, 
                      the tape was moved and a new cycle started. In b) and c), the cycles start with a 10~ms interval without beam. 
           We integrated these time distributions 
           to obtain the number of $^{62}$Ga decays observed. The plots show the complete statistics of the different settings.
}
\label{fig:beta_time}
\end{center}
\end{figure*}

The detection setup consisted of a collection tape (100~$\mu$m thickness of mylar, half an inch wide), a 4$\pi$ cylindrical plastic 
scintillator coupled to two 2-inch photomultipliers (PMs) and 3 HPGe clover detectors from the Euroball array with a relative 
efficiency of 120\%  per detector. This setup is shown schematically in figure~\ref{fig:setup}. The collection tape was controlled 
by stepping motors which enabled a movement of about 10 cm in 100 ms, which was the chosen cycle time for the transport of the tape. 
The event trigger was a coincidence between the two PMs. The $\beta$-detection efficiency was determined with a calibrated $^{90}$Sr 
source to be about 90\%. The $\gamma$ detection efficiency was determined with standard calibration sources ($^{60}$Co, $^{133}$Ba, 
$^{134}$Cs, $^{137}$Cs, and $^{228}$Th). Out of the 12 segments of the three clover detectors, two were rejected
in the final analysis because their energy resolution (typically 3~keV for the good segments) was rather poor ($\approx$10-15~keV).
The final $\gamma$ detection efficiency was about 5.5\% for the 954~keV $\gamma$ ray of $^{62}$Ga and 3.2\% at 2227~keV in the 
add-back mode (see below).

Due to the rather close geometry of the detectors and the large $\beta$-decay $Q$ value, the probability of $\beta$ particles depositing 
energy in the germanium detectors was quite high (about 4\% per crystal). Therefore, corrections to the $\gamma$-ray photopeak efficiency 
had to be applied in order to account for pile-up between a $\gamma$ ray and a $\beta$ particle in the decay of $^{62}$Ga. This 
correction was obtained by means of a Monte-Carlo simulation which included the exact geometry and a realistic $\beta$ spectrum. The 
average correction factor for the branching ratios in singles mode was 3.86(1)\% and 12.85(1)\% for the add-back mode (see below).

The data acquisition system was based on the GANIL data acquisition and allowed an online supervision of the experiment and an 
event-by-event registration of the events. The data were written on DLT tapes for further analysis.

\section{Data analysis}

We analysed the experimental data in two distinct ways: i) by treating the different segments of the clover detectors as 
independent detectors (singles analysis) and ii) by making use of the add-back mode, where we sum the signals from all
crystals of a clover detector provided they were above the noise threshold (100~keV in the present case). Both analyses 
yielded similar results for the branching ratios of all $\gamma$ rays observed. We also analysed the data from the different 
production schemes (with and without JYFLTRAP, grow-in only, grow-in and decay) independently. We obtained consistent results
for all subgroups of our data.

The data obtained on the central beam line, i.e. without the JYFLTRAP system, are contaminated by other A=62 isobars and, to a much 
smaller extent, by A=63 isotopes. In particular, $^{62}$Zn and $^{62}$Cu were strongly produced and transmitted to the detection setup.
The $\gamma$-ray spectra obtained during these measurements were strongly contaminated with $\gamma$ rays from these
isotopes. Therefore, we analysed the data in the following way: The intensities of the 954~keV $\gamma$ ray which de-excites the 
first 2$^+$ state in the $^{62}$Ga $\beta$-decay daughter nucleus $^{62}$Zn and of the 851~keV line (2$^+_2 \rightarrow $ 2$^+_1$)
were determined directly from the $\beta$-gated $\gamma$ spectrum. The other three $\gamma$ rays at 1388~keV, 1850~keV and at 2227~keV, 
although to some extent visible also in the $\beta$-gated spectrum, were analysed in $\beta\gamma$-gated 
spectra, where the $\gamma$ gate was the 954~keV $\gamma$ ray.

This procedure prevents us from observing decays which by-pass the 954~keV level in $^{62}$Zn. However, the background 
in the ungated spectrum was in any case too high to observe directly any other rather weak $\gamma$ ray, not passing through
the 954~keV level.

To determine absolute branching ratios, the source strength has to be known. The number of $^{62}$Ga ions accumulated during 
the different measurement cycles was determined by fitting the time distribution of the $\beta$ events registered with the 
plastic scintillator. As all events were triggered by this plastic scintillator, its own\ $\beta$ detection efficiency is not needed.

To arrive at the final results, we averaged the results obtained by the different analysis procedures (singles and add-back).
The error was determined by averaging the statistical error for each analysis and by adding quadratically the difference
between the average value and the individual results.

\begin{figure*}
\begin{center}
\begin{minipage}{18cm}
\includegraphics[scale=0.29]{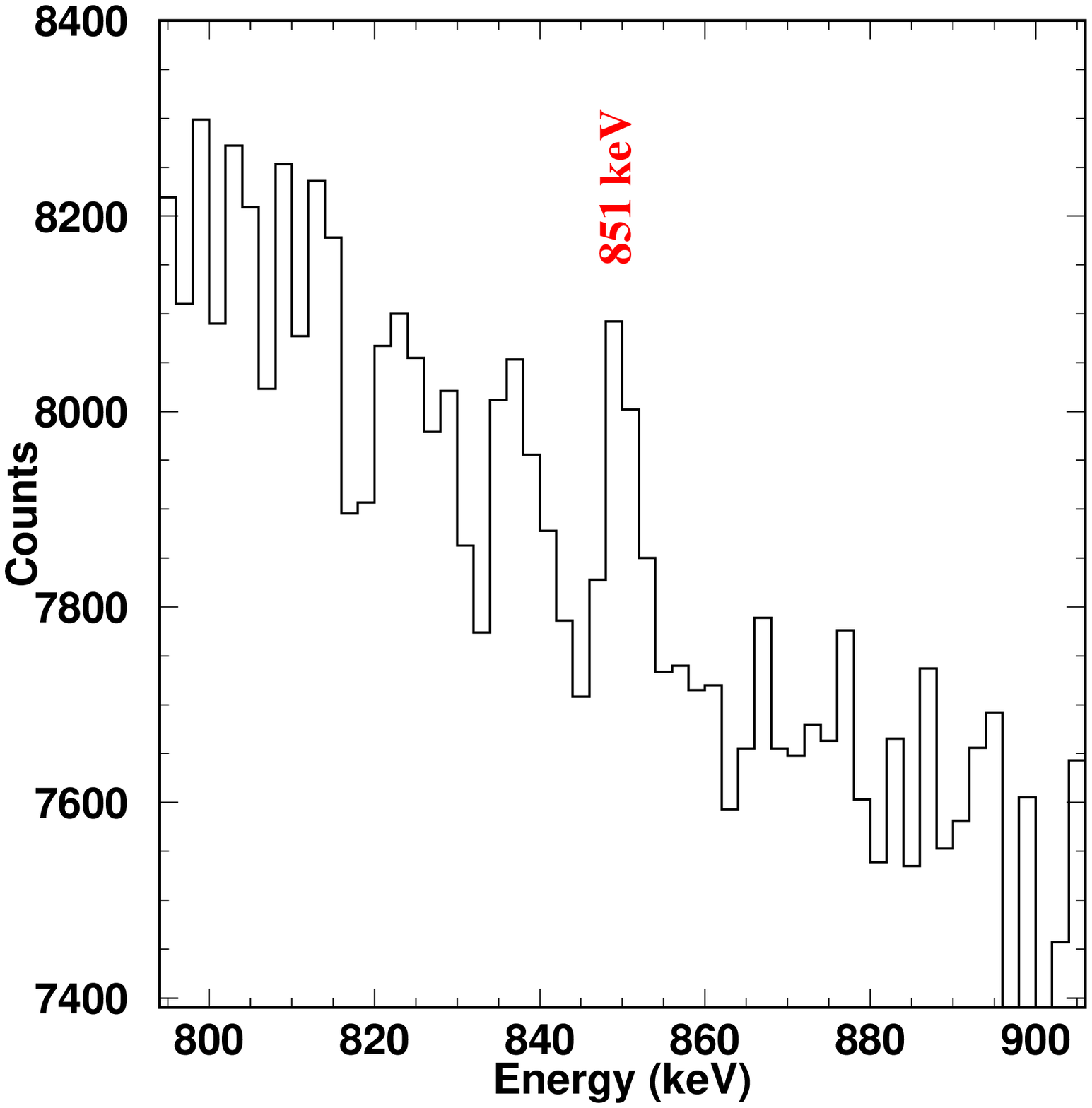} \ \hfill \
\includegraphics[scale=0.29]{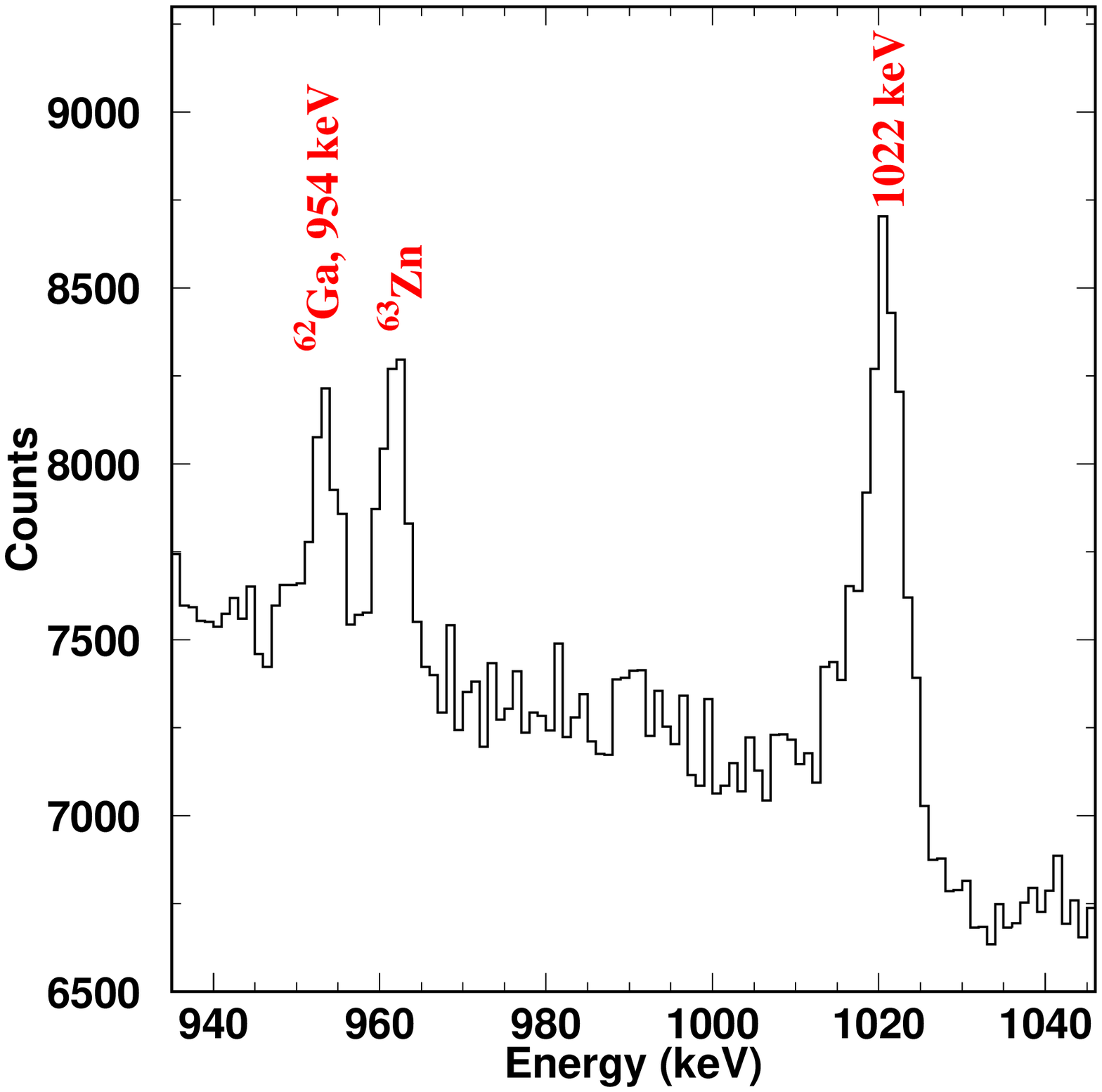} \ \hfill \
\includegraphics[scale=0.29]{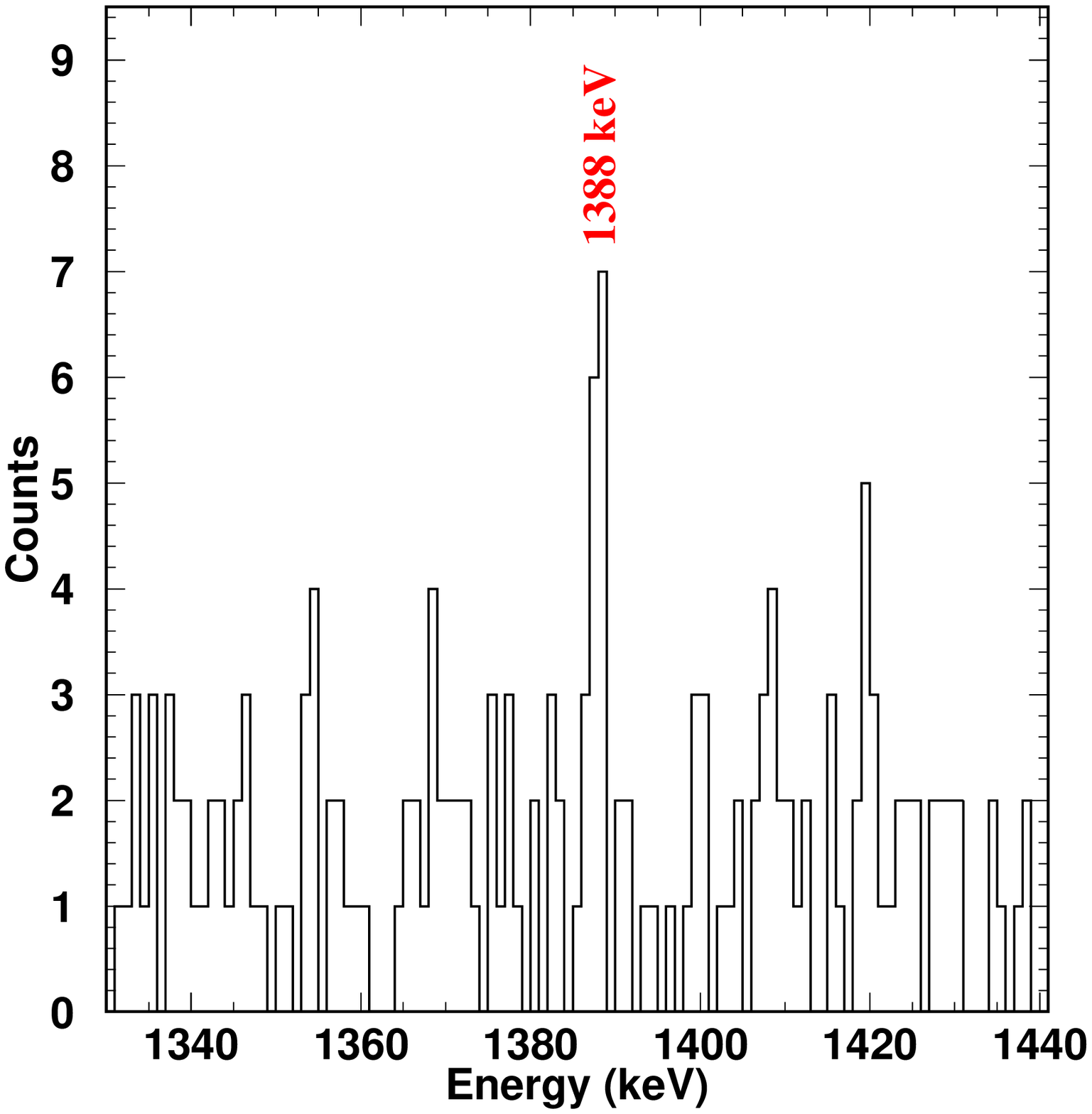} \\
\includegraphics[scale=0.29]{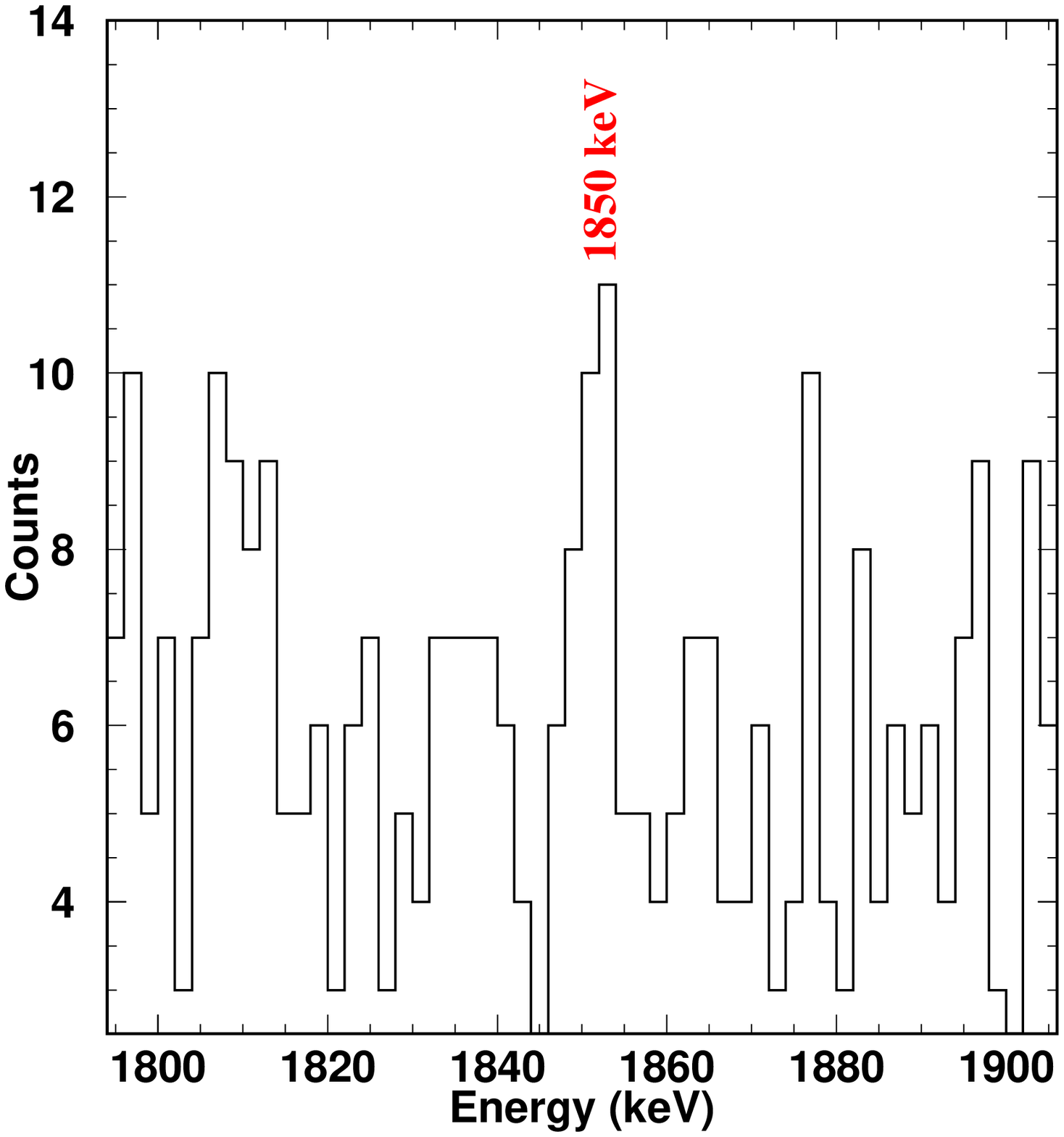} \ \hfill \ 
\includegraphics[scale=0.29]{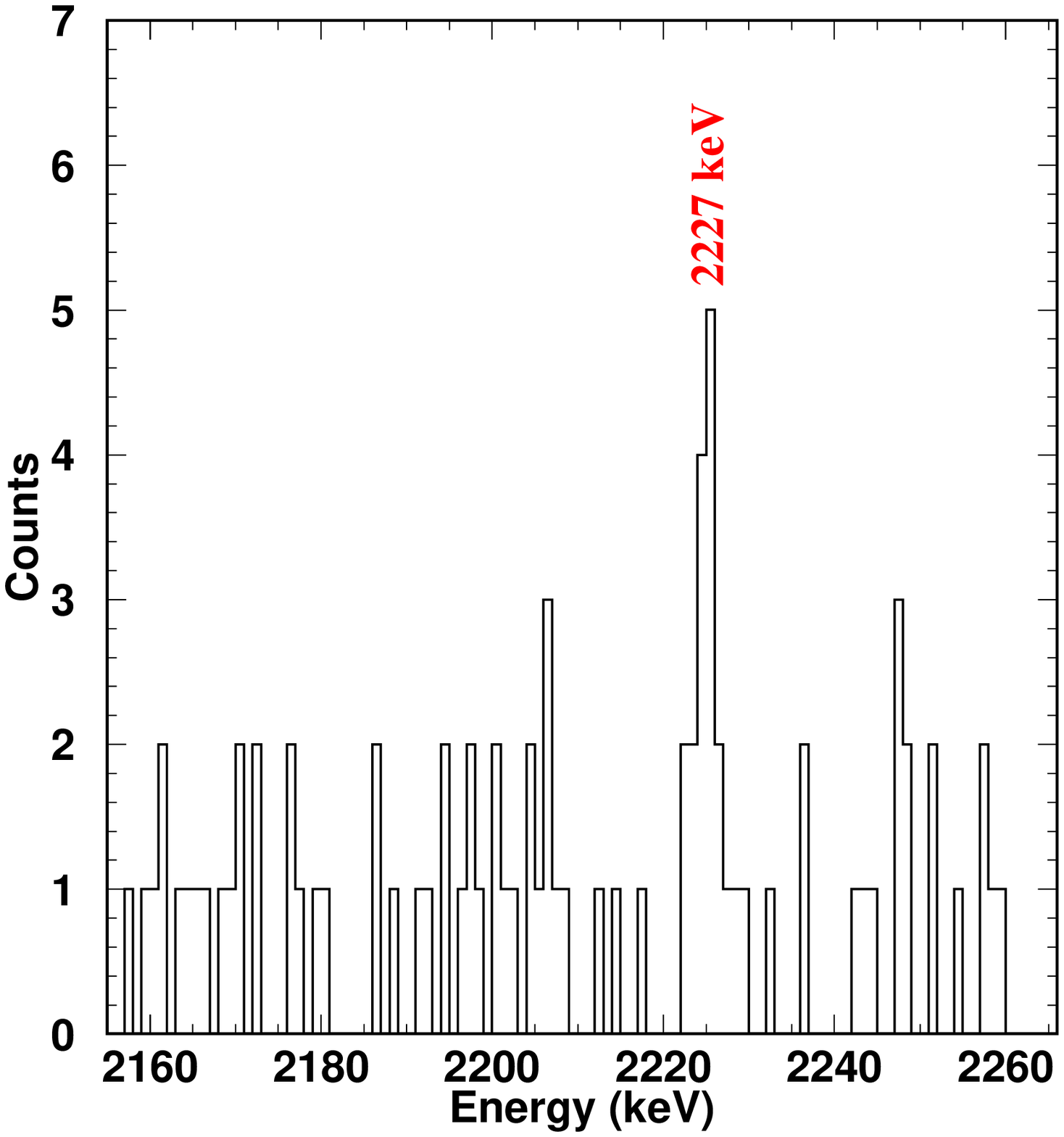} \ \hfill \ \hspace*{5.8cm}
\end{minipage}
\caption[]{Gamma-ray spectra in add-back mode (singles mode for the 1850~keV spectrum) showing regions of specific interest. 
           The spectrum around the peaks at 851~keV and 954~keV 
           are obtained only with a $\beta$-particle coincidence. The other spectra were taken in coincidence with 
           the observation of a 954~keV $\gamma$ ray. As the quality of the spectrum for the 1850~keV line obtained in singles mode is
           much superior to the spectrum obtained in the add-back mode, the singles-mode spectrum is presented here.
}
\label{fig:gammas}
\end{center}
\end{figure*}

\section{Experimental results and comparison with results from the literature}

Figure~\ref{fig:beta_time} shows the time distribution for the three different measurement cycles: 
a) the trap-assisted measurement with the Penning trap system JYFLTRAP, 
b) the measurement on the central IGISOL beam line with the grow-in and decay cycle, and
c) the mesurement on the central beam line where we used only the grow-in cycle. 
The integrated numbers of $^{62}$Ga $\beta$ decays observed during the different cycles are 4.3243(61)$\times$10$^{7}$, 
2.546(18)$\times$10$^{6}$, and 2.4972(30)$\times$10$^{7}$, respectively. To arrive at these results, the time distributions
were fitted with the decay curve of $^{62}$Ga, the grow-in and decay part, and the grow-in part only, respectively.
For each fit, a constant was added for the background. The contribution of the contaminants and of the long-lived daughter nuclei 
was included in the constant background. Their explicit inclusion did not improve the fit.

As mentioned, the $\gamma$-ray branching ratios were determined by means of the singles and the add-back modes. Figure~\ref{fig:gammas} 
shows the parts of the total $\gamma$-ray spectrum, where $\gamma$ rays from the decay of $^{62}$Ga were observed. The central upper 
figure shows the $\gamma$ line due to the de-excitation of the first excited 2$^+$ state in $^{62}$Zn.
Using the total number of counts corrected for the $\gamma$ detection efficiency at 954~keV
and the $\beta$ pile-up, we obtain a branching ratio for this $\gamma$ ray of 0.086(9)\%. For comparison, we give here  
the branching ratio we obtained in the singles mode (0.081(7)\%) and the add-back mode (0.091(8)\%). 
In table~\ref{tab:gamma}, we compare this result with branching ratios for this $\gamma$ ray as found in the literature.

Four other $\gamma$ rays already observed in the work of Hyland et al.~\cite{hyland06} 
were identified in this work.
They appear in the $\beta$-gated $\gamma$ spectra (851~keV line) or in the spectrum additionally conditioned by 
the observation of a 954~keV $\gamma$ ray and are shown in figure~\ref{fig:gammas}.
The branching ratios determined for these $\gamma$ decays are given in table~\ref{tab:gamma}. Their branching ratios compare
reasonably well with the data obtained by Hyland et al.~\cite{hyland06}.

\begin{table*}[hht]
\begin{center}
\begin{tabular}{|c|ccccc|c|}
\hline  \rule{0pt}{1.3em}
energy(keV) & Blank~\cite{blank02ga62} & D{\"o}ring et al.~\cite{doering02} & Hyman et al.~\cite{hyman03} & Canchel et al.~\cite{canchel05} & Hyland et al.~\cite{hyland06} & this work \\
[0.5em] \hline \rule{0pt}{1.3em}
~954        &       0.12(3)\%             &          0.106(17)\%            &         0.120(21)\%         &      0.11(4) \%                 &        0.0809(33)\%           & 0.086(9)\% \\
~851        &           -                 &             -                   &             -               &           -                     &        0.0090(14)\%           & 0.021(8)\% \\
1388        &           -                 &             -                   &             -               &           -                     &        0.0176(20)\%           & 0.023(11)\% \\
1850        &           -                 &             -                   &             -               &           -                     &        0.0081(14)\%           & 0.020(9)\% \\
2227        &           -                 &             -                   &             -               &           -                     &        0.0279(24)\%           & 0.024(10)\% \\
[0.5em] \hline
\end{tabular}
\label{tab:gamma}
\caption{Absolute $\gamma$-ray branching ratios obtained in the present work are compared to values from the literature. For the work of
         Hyland et al.~\cite{hyland06}, we show only the branching ratios of the $\gamma$ rays also observed in the present work.}
\end{center}
\end{table*}

The result of interest is the branching ratio for the super-allowed ground-state to ground-state decay of $^{62}$Ga.
It is evident from the results presented in table~\ref{tab:gamma} that this branching ratio is of the order of 99.9\%. We will use 
two approaches to determine this branching ratio more precisely.

A first approach is to use the calculated strength~\cite{hardy05,hardy02a,towner} which by-passes the first excited 2$^+$ state from 
a shell-model approach. This strength is calculated to be 20\% of the observed decay strength of this 2$^+$ state. If we assume a 
100\% error for this value and take into account that the strength which by-passes the first 2$^+$ state
is certainly not zero, we can adopt a value of 20$^{+20}_{-10}$ \% for this strength. When added to the 
observed strength from the 2$^+$ state, we obtain 0.108$^{+0.029}_{-0.017}$\% for all non-analog branches and 
an analog branching ratio of 99.893$^{+0.018}_{-0.029}$\%. With symmetrized uncertainties, our final result for this 
method is therefore 99.887(23)\%. 

As a second approach we follow the prescription of Hyland et al.~\cite{hyland06}, which uses the fact that 2$^+$ states are not
fed directly by the $\beta$ decay of $^{62}$Ga, but are fed from higher-lying 0$^+$ or 1$^+$ state. The missing feeding is therefore
the difference between the observed decay strength of all 2$^+$ states and their feeding from above. From our data, we 
calculate a missing feeding of the first 2$^+$ state of -0.002(21)\%, of the second 2$^+$ state of 0.021(8)\%, and of 
the third 2$^+$ state of 0.020(9)\%  yielding a total missing feeding of 0.039(38)\%. The strength which by-passes these 
three 2$^+$ states is, according to shell-model calculations~\cite{hardy05,hardy02a,towner}, about 20\% of the branching 
ratio of the 954~keV $\gamma$ ray de-exciting the first 2$^+$ state. If we assume, as above, a 
100\% error for this value and take into account that the strength which by-passes the 2$^+$ states is certainly not 
zero, we can again adopt a value of 20$^{+20}_{-10}$ \% for this strength. With these values, we obtain the unobserved $\gamma$ 
flux to the ground state of 0.010$^{+0.018}_{-0.009}$\%. Combined with the observed $\gamma$ flux via 
the first excited state of 0.086(9)\%, we obtain 0.096$^{+0.028}_{-0.019}$\% for the total 
non-analog strength and thus an analog branch of 99.904$^{+0.019}_{-0.028}$\% or, with symmetric error bars,
99.900(23)\%.

\begin{figure}
\begin{center}
\resizebox{.52\textwidth}{!}{\includegraphics[angle=-90]{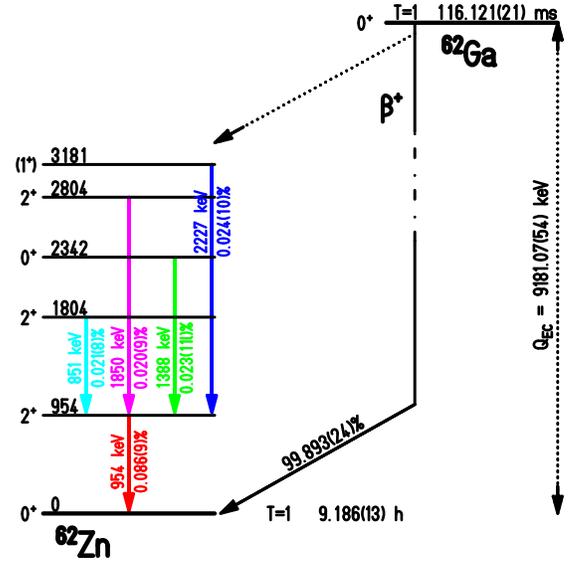}}
\caption[]{Decay scheme of $^{62}$Ga with the $\gamma$ rays and their intensities as determined in the present work.
           Indicated in the figure is also the total Fermi-decay branching ratio, the $^{62}$Ga half-life and the 
           $\beta$-decay $Q$ value.
}
\label{fig:decay_scheme}
\end{center}
\end{figure}

This second result is in excellent agreement with the first one and we adopt a final super-allowed branching ratio of $BR$~= 99.893(24)\%.
Our experimental results are summarised in figure~\ref{fig:decay_scheme}. 

Our value for the super-allowed branching ratio agrees reasonably well with the result obtained by 
Hyland et al.~\cite{hyland06} of 99.861(11)\%. The question is now how to
average these two results to arrive at the final recommended value for the super-allowed branch. The problem is that in both
determinations, in the present work as well as in the work of Hyland et al., the same shell-model calculations have been used 
to obtain the missing strength. We believe, however, that due to the fact that in both estimations a 100\% error was assumed for these
calculations, we can nonetheless average them. We finally obtain an average value of 99.867(10)\%.

Another way of averaging the results from Hyland et al. and our results would be to average the $\gamma$ branching ratios and
then use these averages to calculate the total non-analog branching ratios according to the procedure proposed by Hyland et al.
When we do so, we get a total missing strength for the 2$^+$ states of 0.025(7)\% which has to be compared to the value of Hyland et al.
of 0.024(6)\%. Evidently, this yields the same final result for the non-analog and therefore also for the analog branching ratio.
We prefer the procedure described in the previous paragraph, as it yields independent experimental final results which rely only on
the same theoretical calculation. We therefore keep the final value for the analog branching ratio of 99.867(10)\%.

With the half-life of 116.121(21)~ms~\cite{grinyer08} as well as the Q value of 9181.07(54)~keV~\cite{eronen06} and the statistical rate 
function of 26401.6(83), we obtain an $ft$ value of 3074.1(12)~s which includes the electron-capture correction. 
Using the correction factors as determined by Towner and Hardy~\cite{towner07}, we obtain a $\mathcal{F}t $ value of 3071.4(72)~s.
This value compares well with the most recent evaluation~\cite{towner07} for the average $\mathcal{F}t $ value which yielded 3071.4(8)~s
and which included an $\mathcal{F}t $ value for $^{62}$Ga.

\section{Summary}

We have determined the non-analog $\beta$-decay branching ratios of $^{62}$Ga. The present experimental results together
with shell-model calculations allowed the determination of the super-allo\-wed analog branching ratio for the 0$^+$ to 0$^+$ ground-state 
to ground-state decay to be 99.893(24)\%. The present result, although less precise, is in agreement with the high-precision study 
of Hyland et al. and enables the calculation of an error-weighted average value. Using published values for the $\beta$-decay 
half-life and the $Q$ value, we determine a new $ft$ value of 3074.1(12)~s and a corrected $\mathcal{F}t $ value of 3071.4(72)~s. 
The present value compares well with the average $\mathcal{F}t $ value obtained from 12 other nuclei.

\section*{Acknowledgment}

The authors would like to acknowledge the continous effort of the whole 
Jyv\"askyl\"a accelerator laboratory staff for ensuring a smooth running 
of the experiment. This work was supported in part by the Conseil R\'egional d'Aquitaine
and by the European Union's Sixth Framework Programme "Integrated Infrastructure
Initiative - Transnational Access", Contract Number 506065 (EURONS).
We also acknowledge support from the Academy of Finland under the Finnish
Centre of Excellence Programme 2000-2005 (Project No. 44875, Nuclear and
Condensed Matter Physics Programme at JYFL). The possibility to use detectors from 
the EXOGAM collaboration is gratefully acknowledged.

\end{document}